\documentclass{desyproc}
\usepackage{graphicx,epsfig,color}
\usepackage{wrapfig,rotating}

\begin{document}


\newcommand{\invfb}{\ensuremath{\mathrm{fb^{-1}}}}
\newcommand{\invpb}{\ensuremath{\mathrm{pb^{-1}}}}
\newcommand{\invnb}{\ensuremath{\mathrm{nb^{-1}}}}
\newcommand{\abt} {\mbox{$|t|$}}
\newcommand{\rhoprim}{\mbox{$\rho^\prime$}}
\newcommand{\tprim}{\mbox{$t^\prime$}}
\newcommand{\rhop}{\mbox{$\rho^\prime$}}
\newcommand{\zet}{\mbox{$\zeta$}}
\newcommand{\rfivecomb}{\mbox{$r^5_{00} + 2 r^5_{11}$}}
\newcommand{\ronecomb}{\mbox{$r^1_{00} + 2 r^1_{11}$}}
\newcommand{\gstarVM} {\mbox{$\gamma^* \ \!p \rightarrow V\ \!Y$}}
\newcommand{\gstarVMel} {\mbox{$\gamma^* \ \!p \rightarrow V\ \!p$}}
\newcommand{\gsrel} {\mbox{$\gamma^* \ \!p \rightarrow \rho \ \!p$}}
\newcommand{\gsrpd} {\mbox{$\gamma^* \ \!p \rightarrow \rho \ \!Y$}}
\newcommand{\gspel} {\mbox{$\gamma^* \ \!p \rightarrow \phi \ \!p$}}
\newcommand{\gsppd} {\mbox{$\gamma^* \ \!p \rightarrow \phi \ \!Y$}}
\newcommand{\gstarp} {\mbox{$\gamma^*\ \!p$}}
\newcommand{\mv} {\mbox{$M_V$}}
\newcommand{\mvsq} {\mbox{$M_V^2$}}
\newcommand{\msq} {\mbox{$M_V^2$}}
\newcommand{\qsqplmsq} {\mbox{($Q^2 \!+ \!M_V^2$})}
\newcommand{\qqsqplmsq} {\mbox{$Q^2 \!+ \!M_V^2}$}
\newcommand{\alprim}{\mbox{$\alpha^\prime$}}
\newcommand{\alphaz}{\mbox{$\alpha(0)$}}
\newcommand{\alpomz}{\mbox{$\alpha_{\PO}(0)$}}
\newcommand{\hence}{\mbox{$=>$}}
\newcommand{\vm}{\mbox{$V\!M$}}
\newcommand{\sur}{\mbox{\ \! / \ \!}}
\newcommand{\tzz} {\mbox{$T_{00}$}}
\newcommand{\tuu} {\mbox{$T_{11}$}}
\newcommand{\tzu} {\mbox{$T_{01}$}}
\newcommand{\tuz} {\mbox{$T_{10}$}}
\newcommand{\tmuu} {\mbox{$T_{-11}$}}
\newcommand{\tumu} {\mbox{$T_{1-1}$}}
\newcommand{\ralpha} {\mbox{$\tuu / \tzz$}}
\newcommand{\rbeta} {\mbox{$\tzu / \tzz$}}
\newcommand{\rdelta} {\mbox{$\tuz / \tzz$}}
\newcommand{\reta} {\mbox{$\tmuu / \tzz$}}
\newcommand{\abstzz} {\mbox{$|T_{00}|$}}
\newcommand{\abstuu} {\mbox{$|T_{11}|$}}
\newcommand{\abstzu} {\mbox{$|T_{01}|$}}
\newcommand{\abstuz} {\mbox{$|T_{10}|$}}
\newcommand{\abstmuu} {\mbox{$|T_{-11}|$}}
\newcommand{\rralpha} {\mbox{\abstuu \sur \abstzz}}
\newcommand{\rrbeta} {\mbox{\abstzu \sur \abstzz}}
\newcommand{\rrdelta} {\mbox{\tuz \sur \tzz}}
\newcommand{\rreta} {\mbox{\abstmuu \sur \abstzz}}
\newcommand{\averm} {\mbox{$\av {M}$}}
\newcommand{\rapproch} {\mbox{$R_{SCHC+T_{01}}$}}
\newcommand{\chisq} {\mbox{$\chi^2 / {\rm d.o.f.}$}}

\newcommand{\scaleqsqplmsq} {\mbox{$\qsqplmsq /4$}}


%
%
\newcommand{\s}{\mbox{$s$}}
\newcommand{\ttra}{\mbox{$t$}}
\newcommand{\modt}{\mbox{$|t|$}}
\newcommand{\eminpz}{\mbox{$E-p_z$}}
\newcommand{\eminpzs}{\mbox{$\Sigma(E-p_z)$}}
\newcommand{\rap}{\ensuremath{\eta^*} }
\newcommand{\W}{\mbox{$W$}}
\newcommand{\w}{\mbox{$W$}}
\newcommand{\Q}{\mbox{$Q$}}
\newcommand{\q}{\mbox{$Q$}}
\newcommand{\xB}{\mbox{$x$}}  
\newcommand{\xF}{\mbox{$x_F$}}  
\newcommand{\xg}{\mbox{$x_g$}}  
\newcommand{\xbj}{x}
\newcommand{\xpom}{x_{\PO}}
\newcommand{\y}{\mbox{$y~$}}
\newcommand{\Qsq}{\mbox{$Q^2$}}
\newcommand{\qsq}{\mbox{$Q^2$}}
\newcommand{\kjet}{\mbox{$k_{T\rm{jet}}$}}
\newcommand{\xjet}{\mbox{$x_{\rm{jet}}$}}
\newcommand{\Ejet}{\mbox{$E_{\rm{jet}}$}}
\newcommand{\thjet}{\mbox{$\theta_{\rm{jet}}$}}
\newcommand{\pjet}{\mbox{$p_{T\rm{jet}}$}}
\newcommand{\et}{\mbox{$E_T~$}}
\newcommand{\kt}{\mbox{$k_T~$}}
\newcommand{\ptrans}{\mbox{$p_T~$}}
\newcommand{\pth}{\mbox{$p_T^h~$}}
\newcommand{\pte}{\mbox{$p_T^e~$}}
\newcommand{\ptsq}{\mbox{$p_T^{\star 2}~$}}
\newcommand{\as}{\mbox{$\alpha_s~$}}
\newcommand{\ycut}{\mbox{$y_{\rm cut}~$}}
\newcommand{\gx}{\mbox{$g(x_g,Q^2)$~}}
\newcommand{\xpart}{\mbox{$x_{\rm part~}$}}
\newcommand{\mrsdm}{\mbox{${\rm MRSD}^-~$}}
\newcommand{\mrsdmp}{\mbox{${\rm MRSD}^{-'}~$}}
\newcommand{\mrsdn}{\mbox{${\rm MRSD}^0~$}}
\newcommand{\lambdams}{\mbox{$\Lambda_{\rm \bar{MS}}~$}}
%
%
\newcommand{\gp}{\ensuremath{\gamma}p }
\newcommand{\gammasp}{\ensuremath{\gamma}*p }
\newcommand{\gammap}{\ensuremath{\gamma}p }
\newcommand{\gsp}{\ensuremath{\gamma^*}p }
\newcommand{\dsiget}{\ensuremath{{\rm d}\sigma_{ep}/{\rm d}E_t^*} }
\newcommand{\dsigrap}{\ensuremath{{\rm d}\sigma_{ep}/{\rm d}\eta^*} }
\newcommand{\epem}{\mbox{$e^+e^-$}}
\newcommand{\ep}{\mbox{$ep~$}}
\newcommand{\epl}{\mbox{$e^{+}$}}
\newcommand{\emi}{\mbox{$e^{-}$}}
\newcommand{\epm}{\mbox{$e^{\pm}$}}
\newcommand{\se}{section efficace}
\newcommand{\ses}{sections efficaces}
%
%
\newcommand{\phib}{\mbox{$\varphi$}}
\newcommand{\rh}{\mbox{$\rho$}}
\newcommand{\rhz}{\mbox{$\rh^0$}}
\newcommand{\ph}{\mbox{$\phi$}}
\newcommand{\om}{\mbox{$\omega$}}
\newcommand{\jpsi}{\mbox{$J/\psi$}}
\newcommand{\pipi}{\mbox{$\pi^+\pi^-$}}
\newcommand{\pip}{\mbox{$\pi^+$}}
\newcommand{\pim}{\mbox{$\pi^-$}}
\newcommand{\kk}{\mbox{K^+K^-$}}
\newcommand{\bsl}{\mbox{$b$}}
\newcommand{\alp}{\mbox{$\alpha^\prime$}}
\newcommand{\alpom}{\mbox{$\alpha_{\PO}$}}
\newcommand{\alpomp}{\mbox{$\alpha_{\PO}^\prime$}}
\newcommand{\rzzzz}{\mbox{$r_{00}^{04}$}}
\newcommand{\rzqzz}{\mbox{$r_{00}^{04}$}}
\newcommand{\rzquz}{\mbox{$r_{10}^{04}$}}
\newcommand{\rzqumu}{\mbox{$r_{1-1}^{04}$}}
\newcommand{\ruuu}{\mbox{$r_{11}^{1}$}}
\newcommand{\ruzz}{\mbox{$r_{00}^{1}$}}
\newcommand{\ruuz}{\mbox{$r_{10}^{1}$}}
\newcommand{\ruumu}{\mbox{$r_{1-1}^{1}$}}
\newcommand{\rduz}{\mbox{$r_{10}^{2}$}}
\newcommand{\rdumu}{\mbox{$r_{1-1}^{2}$}}
\newcommand{\rcuu}{\mbox{$r_{11}^{5}$}}
\newcommand{\rczz}{\mbox{$r_{00}^{5}$}}
\newcommand{\rcuz}{\mbox{$r_{10}^{5}$}}
\newcommand{\rcumu}{\mbox{$r_{1-1}^{5}$}}
\newcommand{\rsuz}{\mbox{$r_{10}^{6}$}}
\newcommand{\rsumu}{\mbox{$r_{1-1}^{6}$}}
\newcommand{\rzqik}{\mbox{$r_{ik}^{04}$}}
\newcommand{\rhzik}{\mbox{$\rh_{ik}^{0}$}}
\newcommand{\rhqik}{\mbox{$\rh_{ik}^{4}$}}
\newcommand{\rhaik}{\mbox{$\rh_{ik}^{\alpha}$}}
\newcommand{\rhzzz}{\mbox{$\rh_{00}^{0}$}}
\newcommand{\rhqzz}{\mbox{$\rh_{00}^{4}$}}
\newcommand{\raik}{\mbox{$r_{ik}^{\alpha}$}}
\newcommand{\razz}{\mbox{$r_{00}^{\alpha}$}}
\newcommand{\rauz}{\mbox{$r_{10}^{\alpha}$}}
\newcommand{\raumu}{\mbox{$r_{1-1}^{\alpha}$}}

\newcommand{\R}{\mbox{$R$}}
\newcommand{\rzero}{\mbox{$r_{00}^{04}$}}
\newcommand{\rone}{\mbox{$r_{1-1}^{1}$}}
\newcommand{\costh}{\mbox{$\cos\theta$}}
\newcommand{\cosp}{\mbox{$\cos\psi$}}
\newcommand{\costop}{\mbox{$\cos(2\psi)$}}
\newcommand{\cosd}{\mbox{$\cos\delta$}}
\newcommand{\cossqp}{\mbox{$\cos^2\psi$}}
\newcommand{\cossqt}{\mbox{$\cos^2\theta^*$}}
\newcommand{\sint}{\mbox{$\sin\theta^*$}}
\newcommand{\sintot}{\mbox{$\sin(2\theta^*)$}}
\newcommand{\sinsqt}{\mbox{$\sin^2\theta^*$}}
\newcommand{\costhst}{\mbox{$\cos\theta^*$}}
\newcommand{\vep}{\mbox{$V p$}}
\newcommand{\mpipi}{\mbox{$m_{\pi^+\pi^-}$}}
\newcommand{\mkk}{\mbox{$m_{KK}$}}
\newcommand{\mkaka}{\mbox{$m_{K^+K^-}$}}
\newcommand{\mpp}{\mbox{$m_{\pi\pi}$}}       
\newcommand{\mppsq}{\mbox{$m_{\pi\pi}^2$}}   
\newcommand{\mpi}{\mbox{$m_{\pi}$}}          
\newcommand{\mrho}{\mbox{$m_{\rho}$}}        
\newcommand{\mrhosq}{\mbox{$m_{\rho}^2$}}    
\newcommand{\Gmpp}{\mbox{$\Gamma (\mpp)$}}   
\newcommand{\Gmppsq}{\mbox{$\Gamma^2(\mpp)$}}
\newcommand{\Grho}{\mbox{$\Gamma_{\rho}$}}   
\newcommand{\grho}{\mbox{$\Gamma_{\rho}$}}   
\newcommand{\Grhosq}{\mbox{$\Gamma_{\rho}^2$}}   
%
%
\newcommand{\cm}{\mbox{\rm cm}}
\newcommand{\GeV}{\mbox{\rm GeV}}
\newcommand{\gev}{\mbox{\rm GeV}}
\newcommand{\GeVx}{\rm GeV}
\newcommand{\gevx}{\rm GeV}
\newcommand{\GeVc}{\rm GeV/c}
\newcommand{\gevc}{\rm GeV/c}
\newcommand{\MeVc}{\rm MeV/c}
\newcommand{\mevc}{\rm MeV/c}
\newcommand{\MeV}{\mbox{\rm MeV}}
\newcommand{\mev}{\mbox{\rm MeV}}
\newcommand{\MeVx}{\mbox{\rm MeV}}
\newcommand{\mevx}{\mbox{\rm MeV}}
\newcommand{\GeVsq}{\mbox{${\rm GeV}^2$}}
\newcommand{\gevsq}{\mbox{${\rm GeV}^2$}}
\newcommand{\gevsqc}{\mbox{${\rm GeV^2/c^4}$}}
\newcommand{\gevcsq}{\mbox{${\rm GeV/c^2}$}}
\newcommand{\mevcsq}{\mbox{${\rm MeV/c^2}$}}
\newcommand{\GeVsqm}{\mbox{${\rm GeV}^{-2}$}}
\newcommand{\gevsqm}{\mbox{${\rm GeV}^{-2}$}}
\newcommand{\nb}{\mbox{${\rm nb}$}}
\newcommand{\nbinv}{\mbox{${\rm nb^{-1}}$}}
\newcommand{\pbinv}{\mbox{${\rm pb^{-1}}$}}
\newcommand{\mm}{\mbox{$\cdot 10^{-2}$}}
\newcommand{\mmm}{\mbox{$\cdot 10^{-3}$}}
\newcommand{\mmmm}{\mbox{$\cdot 10^{-4}$}}
\newcommand{\degr}{\mbox{$^{\circ}$}}
%
%
\newcommand{\F}{$ F_{2}(x,Q^2)\,$}  
\newcommand{\Fc}{$ F_{2}\,$}    
\newcommand{\XP}{x_{{I\!\!P}/{p}}}       
\newcommand{\TOSS}{x_{{i}/{\PO}}}        
\newcommand{\un}[1]{\mbox{\rm #1}} 
\newcommand{\LO}{Leading Order}
\newcommand{\NLO}{Next to Leading Order}
\newcommand{\ft}{$ F_{2}\,$}
%
%
\newcommand{\mc}{\multicolumn}
\newcommand{\bce}{\begin{center}}
\newcommand{\ece}{\end{center}}
\newcommand{\beq}{\begin{equation}}
\newcommand{\eeq}{\end{equation}}
\newcommand{\bea}{\begin{eqnarray}}
\newcommand{\eea}{\end{eqnarray}}
%
%
\def\lsim{\mathrel{\rlap{\lower4pt\hbox{\hskip1pt$\sim$}}
    \raise1pt\hbox{$<$}}}         
\def\gsim{\mathrel{\rlap{\lower4pt\hbox{\hskip1pt$\sim$}}
    \raise1pt\hbox{$>$}}}         
%
%
\newcommand{\pom}{{I\!\!P}}
\newcommand{\PO}{I\!\!P}
\newcommand{\slowpi}{\pi_{\mathit{slow}}}
\newcommand{\fiidiii}{F_2^{D(3)}}
\newcommand{\fiidiiiarg}{\fiidiii\,(\beta,\,Q^2,\,x)}
\newcommand{\n}{1.19\pm 0.06 (stat.) \pm0.07 (syst.)}
\newcommand{\nz}{1.30\pm 0.08 (stat.)^{+0.08}_{-0.14} (syst.)}
\newcommand{\fiidiiiful}{F_2^{D(4)}\,(\beta,\,Q^2,\,x,\,t)}
\newcommand{\fiipom}{\tilde F_2^D}
\newcommand{\ALPHA}{1.10\pm0.03 (stat.) \pm0.04 (syst.)}
\newcommand{\ALPHAZ}{1.15\pm0.04 (stat.)^{+0.04}_{-0.07} (syst.)}
\newcommand{\fiipomarg}{\fiipom\,(\beta,\,Q^2)}
\newcommand{\pomflux}{f_{\pom / p}}
\newcommand{\nxpom}{1.19\pm 0.06 (stat.) \pm0.07 (syst.)}
\newcommand {\gapprox}
   {\raisebox{-0.7ex}{$\stackrel {\textstyle>}{\sim}$}}
\newcommand {\lapprox}
   {\raisebox{-0.7ex}{$\stackrel {\textstyle<}{\sim}$}}
\newcommand{\pomfluxarg}{f_{\pom / p}\,(x_\pom)}
\newcommand{\dsf}{\mbox{$F_2^{D(3)}$}}
\newcommand{\dsfva}{\mbox{$F_2^{D(3)}(\beta,Q^2,x_{I\!\!P})$}}
\newcommand{\dsfvb}{\mbox{$F_2^{D(3)}(\beta,Q^2,x)$}}
\newcommand{\dsfpom}{$F_2^{I\!\!P}$}
\newcommand{\gap}{\stackrel{>}{\sim}}
\newcommand{\lap}{\stackrel{<}{\sim}}
\newcommand{\fem}{$F_2^{em}$}
\newcommand{\tsnmp}{$\tilde{\sigma}_{NC}(e^{\mp})$}
\newcommand{\tsnm}{$\tilde{\sigma}_{NC}(e^-)$}
\newcommand{\tsnp}{$\tilde{\sigma}_{NC}(e^+)$}
\newcommand{\st}{$\star$}
\newcommand{\sst}{$\star \star$}
\newcommand{\ssst}{$\star \star \star$}
\newcommand{\sssst}{$\star \star \star \star$}
\newcommand{\tw}{\theta_W}
\newcommand{\sw}{\sin{\theta_W}}
\newcommand{\cw}{\cos{\theta_W}}
\newcommand{\sww}{\sin^2{\theta_W}}
\newcommand{\cww}{\cos^2{\theta_W}}
\newcommand{\trm}{m_{\perp}}
\newcommand{\trp}{p_{\perp}}
\newcommand{\trmm}{m_{\perp}^2}
\newcommand{\trpp}{p_{\perp}^2}
\newcommand{\ev}{\'ev\'enement}
\newcommand{\evs}{\'ev\'enements}
\newcommand{\mdv}{mod\`ele \`a dominance m\'esovectorielle}
\newcommand{\mdmv}{mod\`ele \`a dominance m\'esovectorielle}
\newcommand{\mdm}{mod\`ele \`a dominance m\'esovectorielle}
\newcommand{\idiff}{interaction diffractive}
\newcommand{\idiffs}{interactions diffractives}
\newcommand{\pdmv}{production diffractive de m\'esons vecteurs}
\newcommand{\pdmr}{production diffractive de m\'esons \rh}
\newcommand{\pdmp}{production diffractive de m\'esons \ph}
\newcommand{\pdmo}{production diffractive de m\'esons \om}
\newcommand{\pdm}{production diffractive de m\'esons}
\newcommand{\pdiff}{production diffractive}
\newcommand{\diff}{diffractive}
\newcommand{\produ}{production}
\newcommand{\mvs}{m\'esons vecteurs}
\newcommand{\me}{m\'eson}
\newcommand{\mr}{m\'eson \rh}
\newcommand{\mph}{m\'eson \ph}
\newcommand{\mo}{m\'eson \om}
\newcommand{\mrs}{m\'esons \rh}
\newcommand{\mps}{m\'esons \ph}
\newcommand{\mos}{m\'esons \om}
\newcommand{\photo}{photoproduction}
\newcommand{\agq}{\`a grand \qsq}
\newcommand{\agqsq}{\`a grand \qsq}
\newcommand{\apq}{\`a petit \qsq}
\newcommand{\apqsq}{\`a petit \qsq}
\newcommand{\de}{d\'etecteur}
%
%
\newcommand{\sqrts}{$\sqrt{s}$}
\newcommand{\Oa}{$O(\alpha_s)$}
\newcommand{\Oaa}{$O(\alpha_s^2)$}
\newcommand{\PT}{p_{\perp}}
\newcommand{\sh}{\hat{s}}
\newcommand{\uh}{\hat{u}}
\newcommand{\ttbs}{\char'134}
\newcommand{\xpomlo}{3\times10^{-4}}
\newcommand{\xpomup}{0.05}
\newcommand{\llq}{$\alpha_s \ln{(\qsq / \Lambda_{QCD}^2)}$}
\newcommand{\llqx}{$\alpha_s \ln{(\qsq / \Lambda_{QCD}^2)} \ln{(1/x)}$}
\newcommand{\llx}{$\alpha_s \ln{(1/x)}$}
%
%
\newcommand{\Brodsky}{Brodsky {\it et al.}}
\newcommand{\FKS}{Frankfurt, Koepf and Strikman}
\newcommand{\Kop}{Kopeliovich {\it et al.}}
\newcommand{\Ginzburg}{Ginzburg {\it et al.}}
\newcommand{\Ryskin}{\mbox{Ryskin}}
\newcommand{\Kaidalov}{Kaidalov {\it et al.}}
%
%
\def\ar#1#2#3   {{\em Ann. Rev. Nucl. Part. Sci.} {\bf#1} (#2) #3}
\def\epj#1#2#3  {{\em Eur. Phys. J.} {\bf#1} (#2) #3}
\def\err#1#2#3  {{\it Erratum} {\bf#1} (#2) #3}
\def\ib#1#2#3   {{\it ibid.} {\bf#1} (#2) #3}
\def\ijmp#1#2#3 {{\em Int. J. Mod. Phys.} {\bf#1} (#2) #3}
\def\jetp#1#2#3 {{\em JETP Lett.} {\bf#1} (#2) #3}
\def\mpl#1#2#3  {{\em Mod. Phys. Lett.} {\bf#1} (#2) #3}
\def\nim#1#2#3  {{\em Nucl. Instr. Meth.} {\bf#1} (#2) #3}
\def\nc#1#2#3   {{\em Nuovo Cim.} {\bf#1} (#2) #3}
\def\np#1#2#3   {{\em Nucl. Phys.} {\bf#1} (#2) #3}
\def\pl#1#2#3   {{\em Phys. Lett.} {\bf#1} (#2) #3}
\def\prep#1#2#3 {{\em Phys. Rep.} {\bf#1} (#2) #3}
\def\prev#1#2#3 {{\em Phys. Rev.} {\bf#1} (#2) #3}
\def\prl#1#2#3  {{\em Phys. Rev. Lett.} {\bf#1} (#2) #3}
\def\ptp#1#2#3  {{\em Prog. Th. Phys.} {\bf#1} (#2) #3}
\def\rmp#1#2#3  {{\em Rev. Mod. Phys.} {\bf#1} (#2) #3}
\def\rpp#1#2#3  {{\em Rep. Prog. Phys.} {\bf#1} (#2) #3}
\def\sjnp#1#2#3 {{\em Sov. J. Nucl. Phys.} {\bf#1} (#2) #3}
\def\spj#1#2#3  {{\em Sov. Phys. JEPT} {\bf#1} (#2) #3}
\def\zp#1#2#3   {{\em Zeit. Phys.} {\bf#1} (#2) #3}
%
%
\newcommand{\clearemptydoublepage}{\newpage{\pagestyle{empty}\cleardoublepage}}
\newcommand{\scaption}[1]{\caption{\protect{\footnotesize  #1}}}
\newcommand{\proc}[2]{\mbox{$ #1 \rightarrow #2 $}}
\newcommand{\average}[1]{\mbox{$ \langle #1 \rangle $}}
\newcommand{\av}[1]{\mbox{$ \langle #1 \rangle $}}



\title{Exclusive Hard Diffraction at HERA \\ (DVCS and Vector Mesons) }

\author{{\slshape Pierre Marage}\\[1ex]
Universit\'e Libre de Bruxelles, Boulevard du Triomphe, B-1050 Bruxelles, Belgium \\
{\it On behalf of the H1 and ZEUS Collaborations} }

\contribID{marage\_pierre}

\desyproc{DESY-PROC-2009-xx}
\acronym{EDS'09} 
\doi  

\maketitle

\begin{abstract}
Recent results obtained at HERA on deeply virtual Compton scattering and exclusive vector 
meson production are reviewed, with the emphasis on the transition from soft to hard 
diffraction and on spin dynamics.
\end{abstract}

\section {Introduction}
%
Since the beginning of HERA data taking, a large number of studies have been 
performed of deeply virtual Compton scattering (DVCS) and of vector meson  (VM) production.
The exclusive final states include real photons~\cite{z-dvcs,h1-dvcs,h1-gamma-larget}, 
light ($\rho$~\cite{z-rho-photoprod,h1-rho-photoprod,z-rho,h1-rho-hera1,z-high-t,h1-rho-photoprod-large-t},
  $\omega$~\cite{z-omega-photoprod,z-omega}
  and \ph~\cite{z-phi-photoprod,z-phi,h1-rho-hera1})
and heavy VMs 
(\jpsi~\cite{z-jpsi-photoprod,z-jpsi-elprod,h1-jpsi-hera1,h1-jpsi-large-t,z-jpsi-large-t}, 
$\psi(2s)$~\cite{h1-psi2s} 
and $\Upsilon$~\cite{h1-upsilon,z-upsilon}).
Cross sections are expressed in terms of $\gamma^*p$ scattering.

In the presence of a hard scale, these processes provide unique information 
on the mechanisms of diffraction, in particular on the transition from soft to hard 
diffraction and on spin dynamics.
A hard scale is provided by the VM mass $M_V$, 
by the negative square of the photon four-momentum, \qsq\
(with $\qsq \simeq 0$ for photoproduction and 
$1.5 \leq Q^2 \leq 90~\gevsq$ for electroproduction),
or by the square of the four-momentum transfer at the proton vertex, $t$.

DVCS and VM production with small \modt\ values 
($\modt \leq 0.5~\gevsq$ for elastic scattering and $\modt \leq 2.5~\gevsq$
for proton dissociation)
are interpreted in terms of two complementary QCD approaches.
Following a collinear factorisation theorem, the DVCS process, the electroproduction of 
light VMs by longitudinally polarised photons, and the production of heavy VMs 
can be described by the convolution of the hard process 
with generalised parton distributions in the proton (GPDs).
High energy DVCS andVM production can also be described through the
factorisation of virtual photon fluctuation into a $q \bar q$ colour dipole, 
diffractive dipole--proton scattering, 
and $q \bar q$ recombination into the final state photon or VM.
The interaction scale $\mu$ is given by the characteristic transverse size of 
the dipole, with $\mu^2 \simeq \scaleqsqplmsq$ for light 
VM electroproduction by longitudinal photons and for heavy VM production, 
whereas this value may be significantly reduced 
for light VM electroproduction by transversely polarised photons, 
because of end-point contributions in the photon wave function.
For DVCS, LO contributions are present, which suggests that the relevant scale may 
be \qsq\ rather than $\qsq/4$ as for VMs.

Several models, based on either approach, have been proposed.
They differ in particular in the way the VM wave function is taken into account, 
in the parameterisation of the parton distributions and of the dipole--proton scattering, 
and in the extension to non-zero \modt\ values of the scattering amplitude.

High \modt\ photoproduction, with $2 \leq \modt \leq 30~\gevsq$,
of real photons~\cite{h1-gamma-larget}, 
\rh\ and \ph\ mesons~\cite{z-high-t,h1-rho-photoprod-large-t},
and \jpsi\ mesons~\cite{h1-jpsi-large-t,z-jpsi-large-t} 
offer specific testing grounds for the BFKL evolution.

Comparisons of models with the data are discussed in particular 
in~\cite{z-dvcs,h1-dvcs,h1-gamma-larget,z-rho,h1-rho-hera1,h1-rho-photoprod-large-t,
h1-jpsi-large-t,z-jpsi-large-t,levy}.

\section{Kinematic dependences}
%

\paragraph {{\bf DVCS}}

\begin{figure}[htbp]
\begin{center}
\setlength{\unitlength}{1.0cm}
\begin{picture}(11.,5.)   
\put(0.0,0.0){\epsfig{file=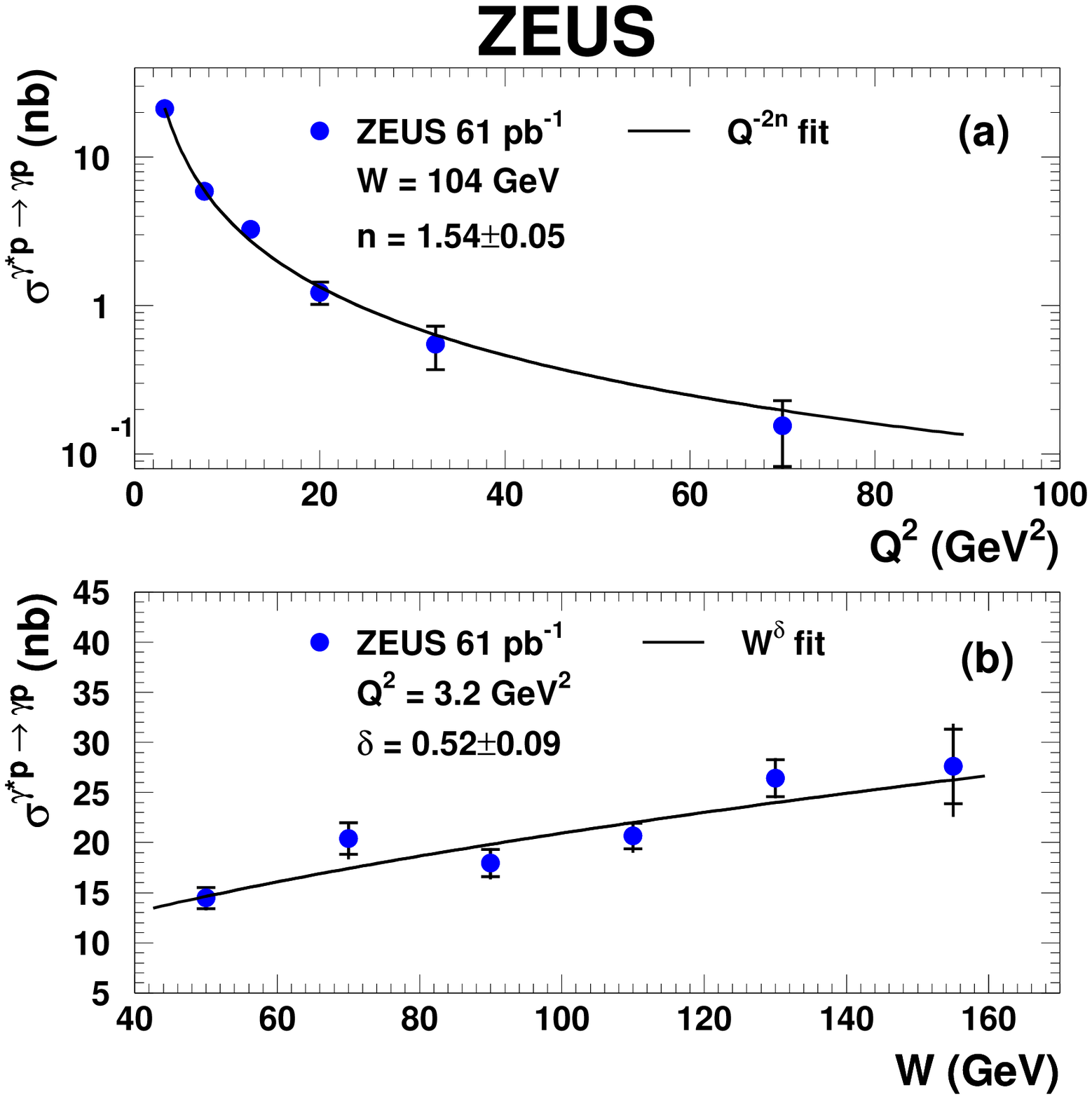,width=6.cm,height=5.cm}}
\put(6.,0.8){\epsfig{file=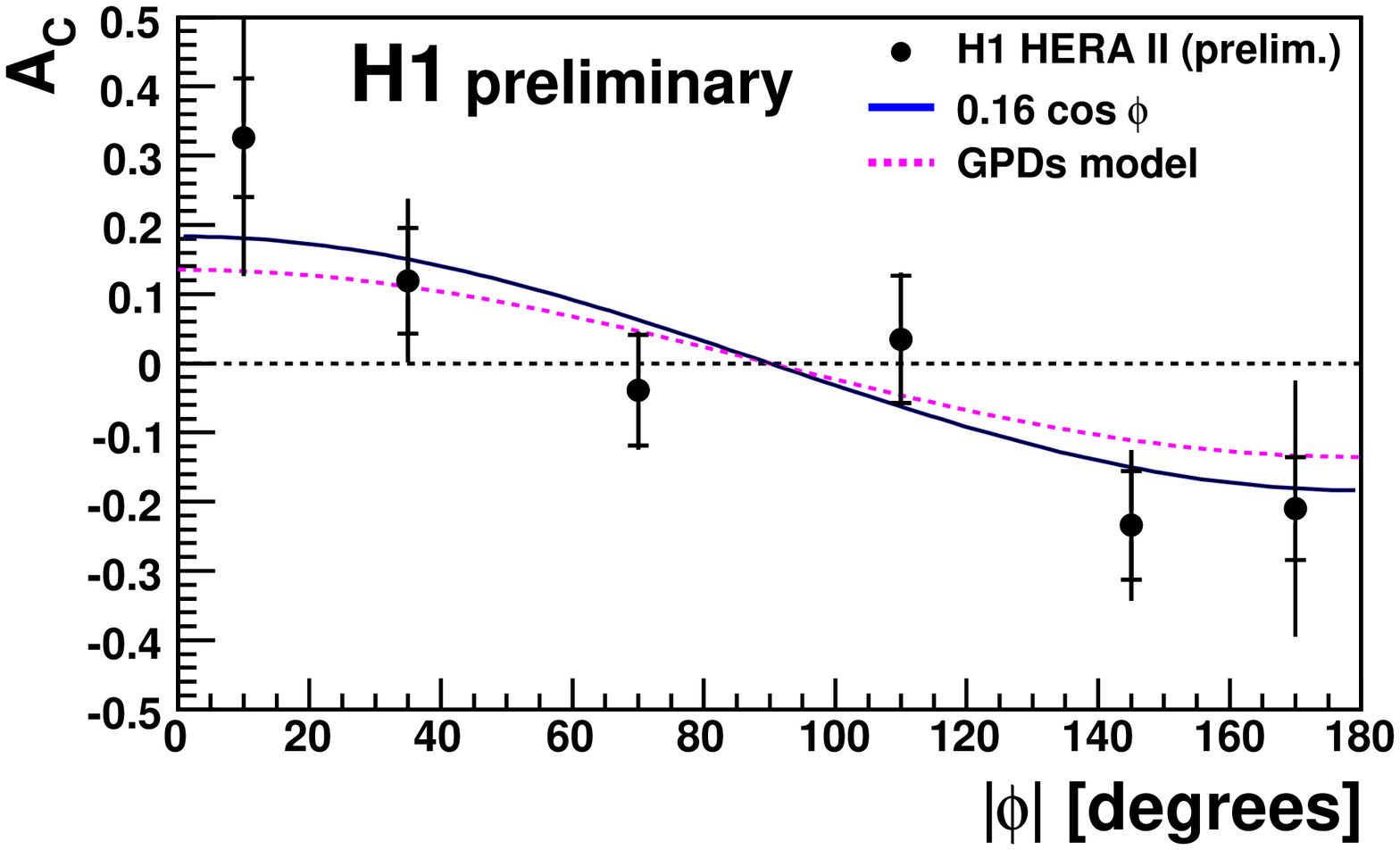,width=5.0cm}}
\end{picture}
\vspace{-0.3cm}
\caption{(left) \qsq\ and $W$ dependences of DVCS production, with simple
fit parameterisations~\protect\cite{z-dvcs};
(right) beam charge asymmetry, $\cos \phi$ fit and predictions of a GPD 
model~\protect\cite{h1-dvcs}.}
\label{fig:dvcs}
\end{center}
\vspace{-0.3cm}
\end{figure}

The kinematic dependences of DVCS production, presented in
Fig.~\ref{fig:dvcs}, are well described by models using either GPDs 
or a dipole approach~\cite{z-dvcs,h1-dvcs}.

The interference of the DVCS and Bethe-Heitler processes gives access, 
through the measurement of beam charge asymmetry, to the ratio $\rho$ of the
real to imaginary parts of the DVCS amplitude.
The measurement $\rho = 0.20 \pm 0.05 \pm 0.08$~\cite{h1-dvcs} is in 
agreement with the value $\rho = 0.25 \pm 0.03 \pm 0.05$ obtained from a 
dispersion relation using the $W$ dependence of the cross section.
%
%

\paragraph {{\bf \qsq\ dependence of light VM production}}

\begin{figure}[h]
\begin{minipage}{0.48\columnwidth}
\centerline{\includegraphics[width=0.80\columnwidth]{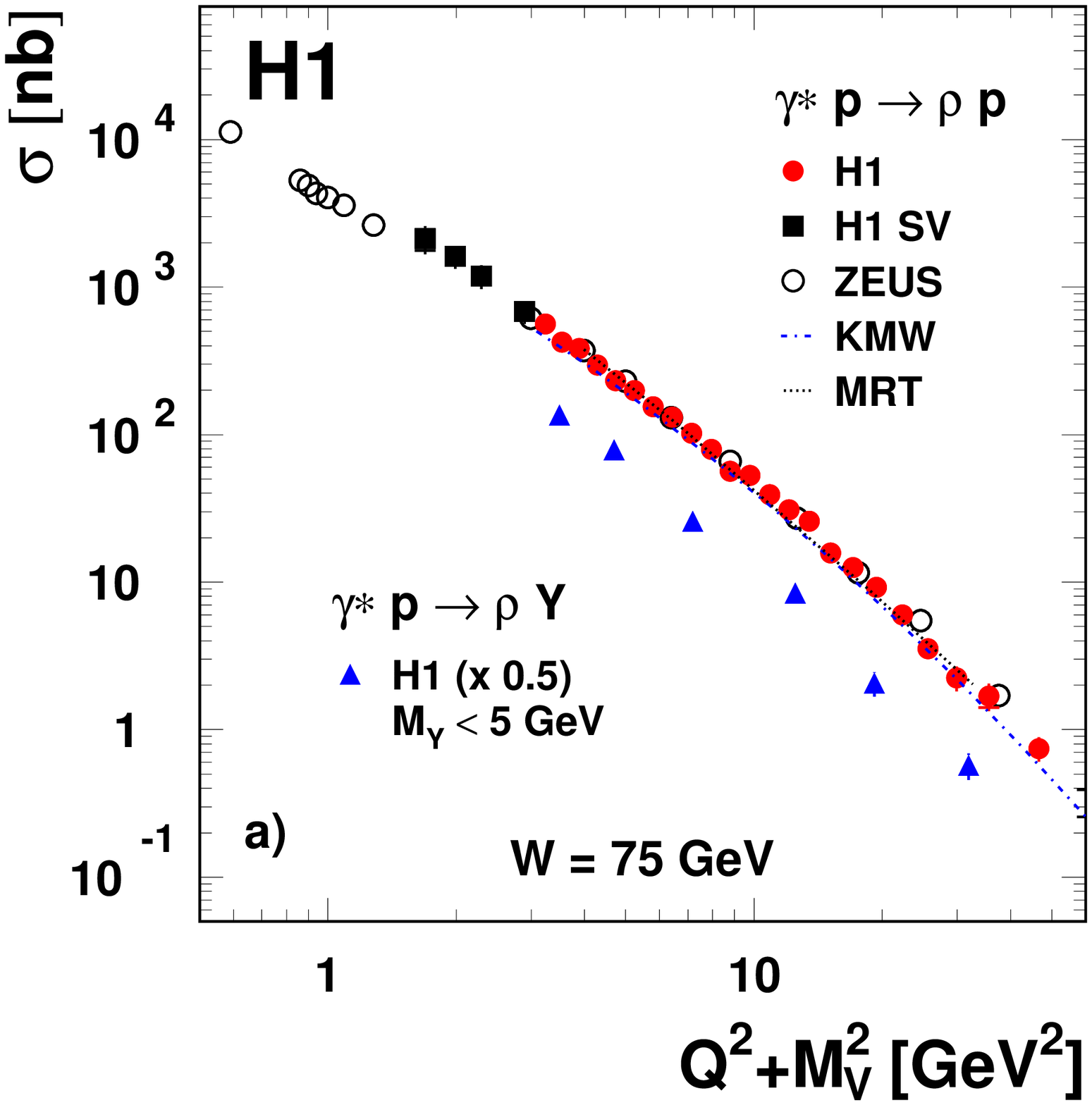}}
\vspace{-0.3cm}
\caption{\qsq\ dependence of elastic and proton dissociative electroproduction 
cross sections of \rh\ mesons, and model predictions~\protect\cite{h1-rho-hera1}.}
\label{fig:rho_q2}
\end{minipage}
\hspace{2mm}
\begin{minipage}[h]{0.48\columnwidth}
\centerline{\includegraphics[width=0.80\columnwidth]{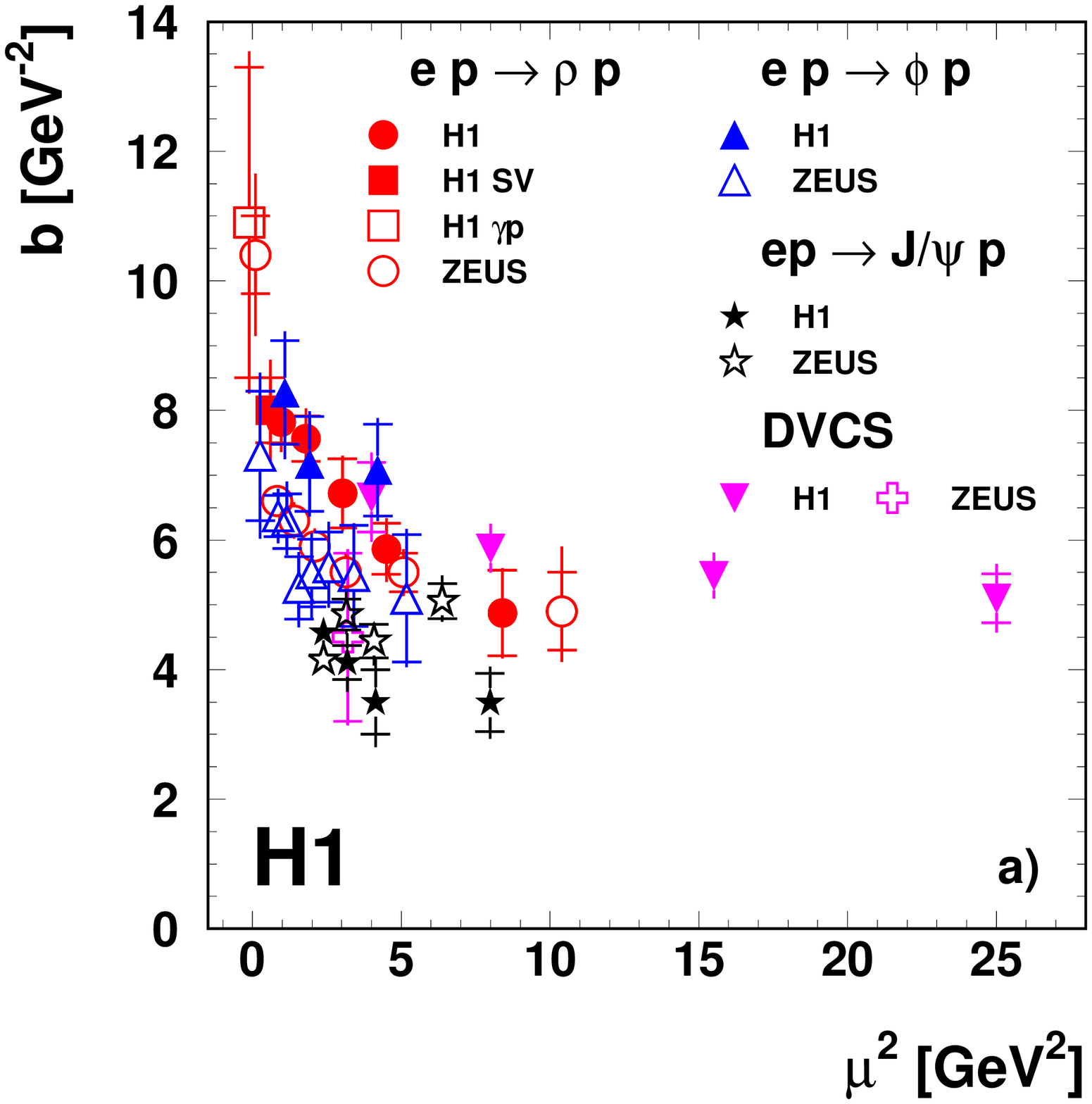}}
\vspace{-0.3cm}
\caption{Elastic $b$ slopes, as a function of $\mu^2 = \scaleqsqplmsq$
for VM production and $\mu^2 = \qsq$ for DVCS~\protect\cite{h1-rho-hera1}.}
\label{fig:b-slopes}
\end{minipage}
\vspace{-0.3cm}
\end{figure}

The cross sections for elastic and proton dissociative \rh\ and \ph\ 
electroproduction have been measured with high 
precision~\cite{z-rho,h1-rho-hera1,z-phi}.
The \qsq\ dependence, shown for \rh\ mesons in Fig.~\ref{fig:rho_q2}, is reasonably 
described by several models, using either the GPD or the dipole approach.

Although the production cross sections for light and heavy VMs differ by several orders 
of magnitude at $\qsq \simeq 0$,
it is striking that the ratios are nearly constant when they are studied as a function of the 
scaling variable \scaleqsqplmsq, with values close to unity when scaled 
according to the quark charge content of the VMs, $\rh : \omega : \phi : \jpsi  = 9 : 1 : 2 : 8$. 
This confirms the relevance of the dipole size to the cross sections, even though the
agreement with SU(4) universality is not perfect, indicating that
wave function effects may need to be taken into account.

The ratio of the production cross sections with proton dissociative and elastic scattering
at $\modt = 0$ is found to be independent of \qsq.
Consistent values around 0.160 are measured for \rh\ and \ph\ production with 
dissociative mass $M_Y < 5~\gev$~\cite{h1-rho-hera1}. 
This observation supports the independence of the hard and soft vertex contributions to the 
scattering amplitudes, known as proton vertex or ``Regge" factorisation.

\paragraph {{\bf $t$ slopes}}

Exponentially falling \modt\ distributions, with $\rm {d} \sigma / \rm {d}t \propto e^{-b |t|}$,
are measured for DVCS, light and heavy VM production, in both the 
elastic and the proton dissociative channels.
In an optical model appraoch, the slope $b$ is given by the convolution of the 
transverse sizes of the
$q \bar q$ dipole, of the diffractively scattered system (which vanishes
for proton dissociation) and of the exchange (a contribution which is expected to be small).
As shown in Fig.~\ref{fig:b-slopes}, the elastic slopes for light VMs
strongly decrease with increasing \qsq.
They reach values of the order of $5~\gevsqm$, similar to those measured in 
\jpsi\ production, for values of the scaling variable
$\scaleqsqplmsq\ \gapprox\ 5~\gevsq$.
This evolution reflects the shrinkage with \qsq\ of the light quark colour dipole.
A similar evolution is observed for DVCS as a function of the variable \qsq.
The proton dissociative slopes similarly decrease with the increasing scale, down to values 
around $1.5~\gevsqm$ for \rh\ production and DVCS, and values slightly below $1~\gevsqm$ 
for \jpsi\ production.

The difference between the elastic and proton dissociative slopes,
$b_{el} - b_{p. diss.}$, provides another test of proton vertex factorisation.
A value of $3.5 \pm 0.1~\gevsqm$ is measured for 
\jpsi~\protect\cite{z-jpsi-photoprod,h1-jpsi-hera1}, with a similar value for 
DVCS~\protect\cite{h1-dvcs}. 
The difference is higher, around $5.5~\gevsqm$, for \rh\ and \ph\ mesons, with
however an indication of a decrease toward the \jpsi\ value with increasing 
$\scaleqsqplmsq$~\cite{h1-rho-hera1}.

\paragraph {{\bf Energy dependence and effective Regge trajectory}}

The energy dependence of DVCS and VM production
is well described by a power law, $\rm {d} \sigma / \rm {d}W \propto W^{\delta}$.

\begin{figure}[h]
\begin{center}
\setlength{\unitlength}{1.0cm}
\begin{picture}(14.,5.5)   
\put(0.0,0.){\epsfig{file=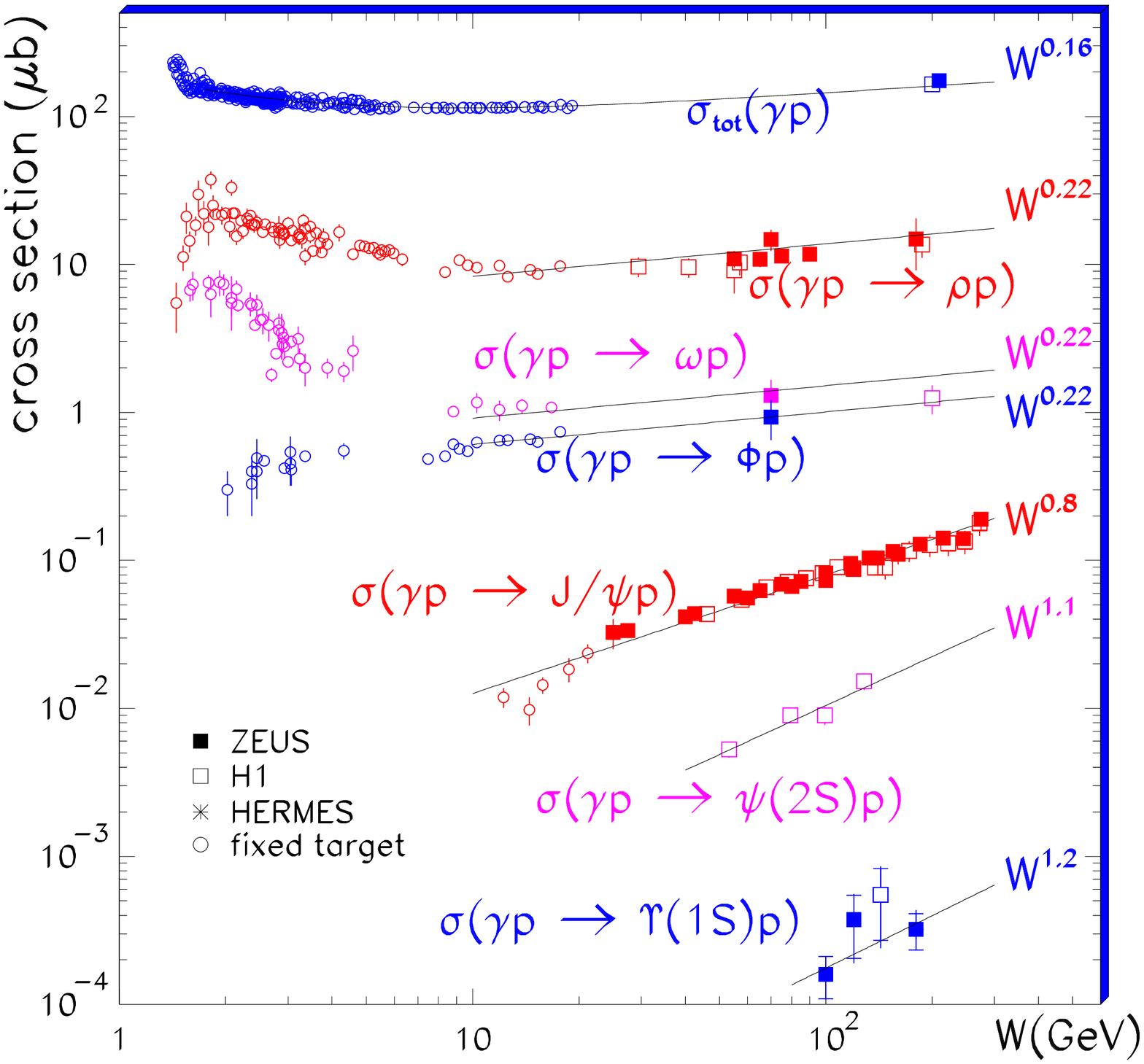,width=5.5cm}}
\put(6.,0.5){\epsfig{file=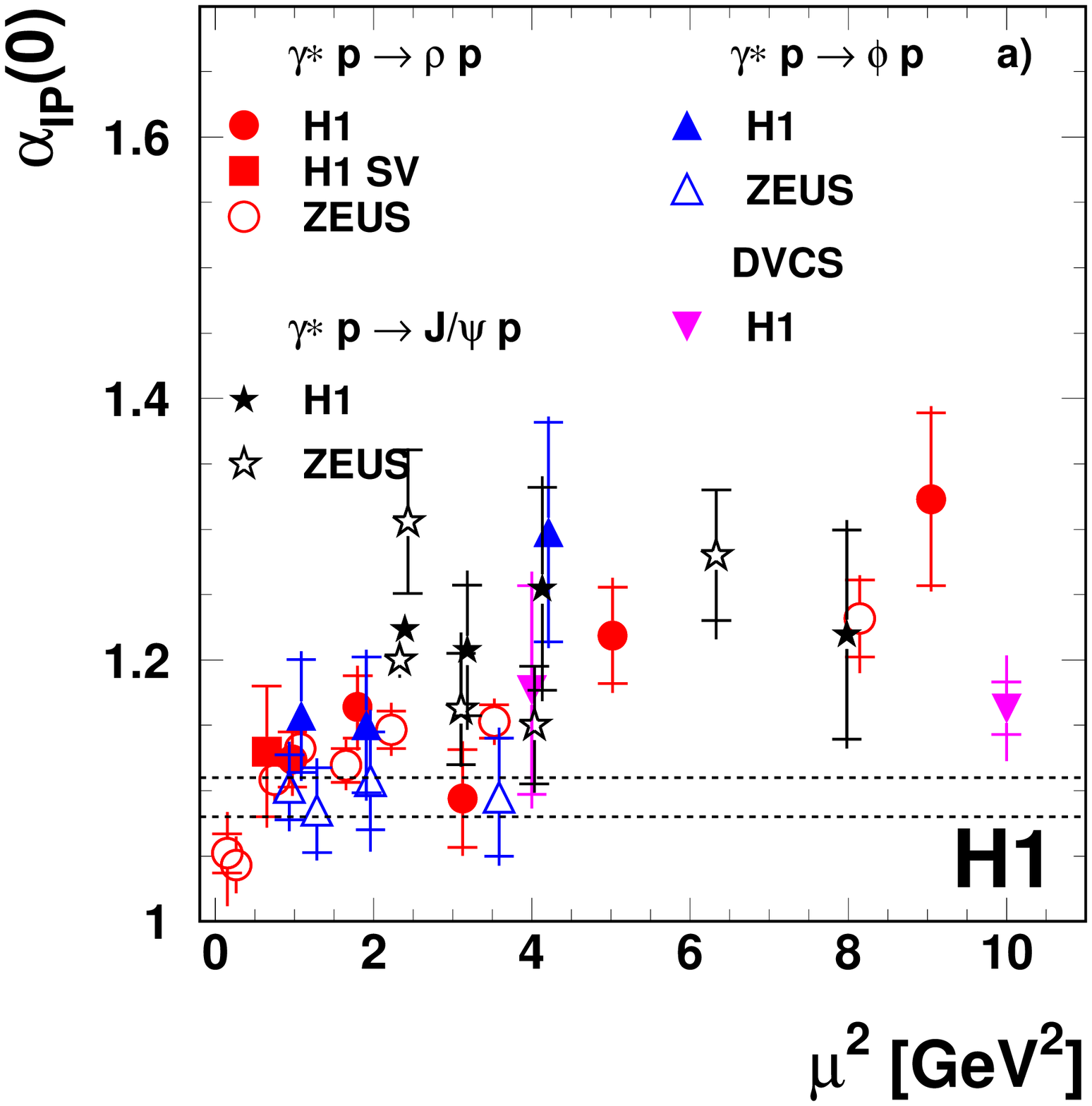,width=4.0cm}}
\put(10.,0.5){\epsfig{file=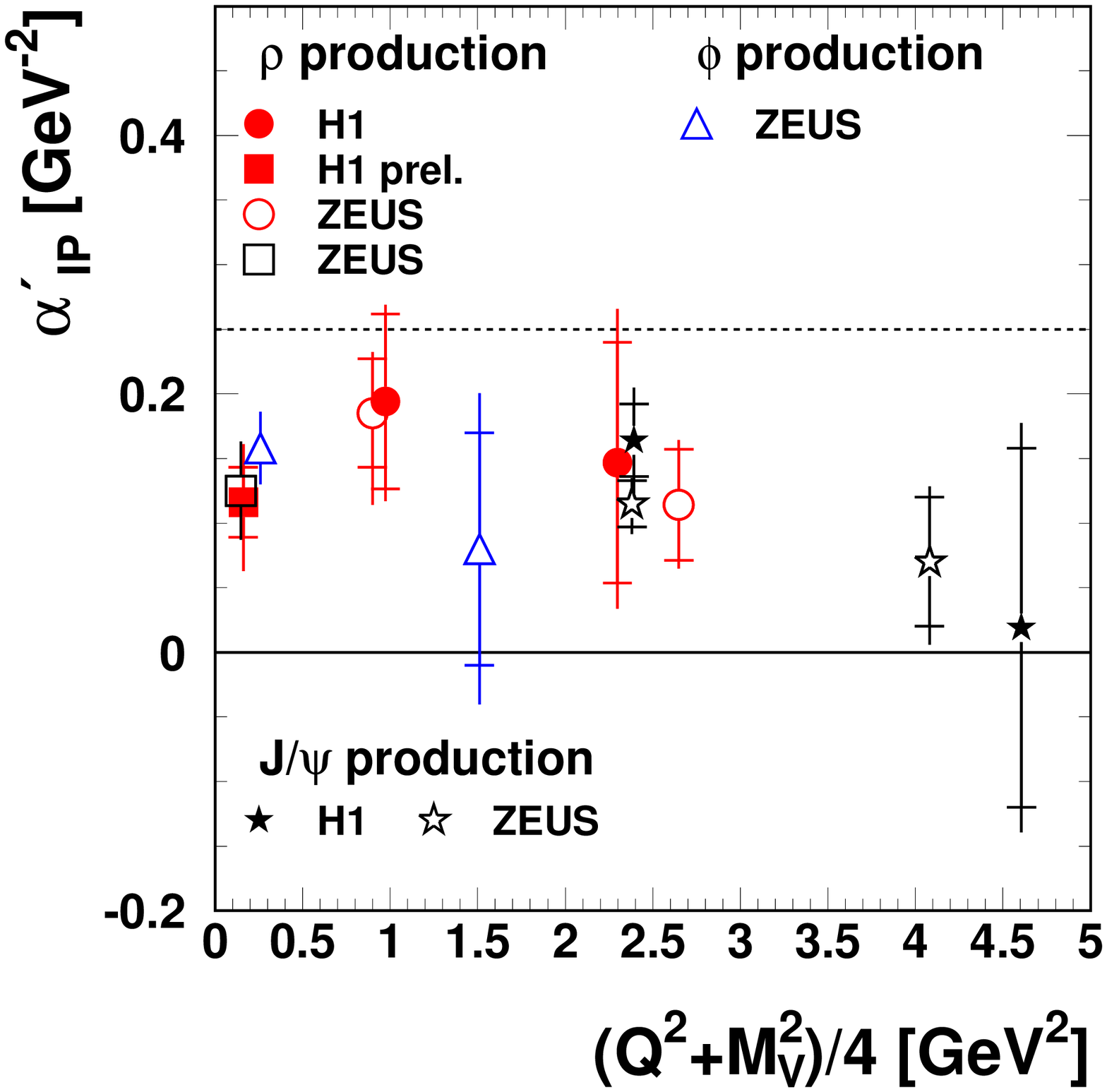,width=4.0cm}}
\end{picture}
\vspace{-0.3cm}
\caption{(left) $W$ dependence of VM photoproduction~\protect\cite{levy}; 
measurement of  the intercept $\alpom(0)$ (centre) and of  the slope \alp\
(right) of the effective Regge trajectory, as a function of the scale 
$\mu^2 =  \qsqplmsq /4$ for VM production and $\mu^2 = \qsq$ 
for DVCS~\protect\cite{h1-rho-hera1}.}
\label{fig:Regge}
\end{center}
\vspace{-0.5cm}
\end{figure}

Figure~\ref{fig:Regge}~(left) shows that the energy dependence is significantly stronger
for heavy quark photoproduction, with $\delta \sim 0.8-1.2$, than for (soft) 
hadron--hadron interactions and light VM photoproduction, with $\delta \sim 0.2$.
This is explained by the fact that the photoproduction of VMs formed of heavy quarks
is a hard process, characterised by small transverse dipoles which probe the low-$x$ 
gluon density in the proton at a scale where it is quickly increasing with $1/x$. 

For light VM production, the $W$ dependence is hardening with \qsq, with values of 
$\delta$ similar to the \jpsi\ values for $\scaleqsqplmsq\ \gapprox\ 5~\gevsq$.
This feature is explained by the shrinkage of the colour dipoles at large \qsq\ 
values.

In a Regge inspired parameterisation, the energy dependence of the
cross section and its correlation with $t$ are given by
$\delta(t) =  4 \ ( \alpom(t)  - 1)$, with $\alpom(t) = \alpom(0) + \alp \cdot \ t$,
where \alp\ describes the shrinking of the diffractive peak with energy.
The hard behaviour of \jpsi\ production and the hardening with $\scaleqsqplmsq$ of 
the energy dependence of light VM production for $t = 0$ are shown in 
Fig.~\ref{fig:Regge}~(centre), where values of $1.08$ or $1.11$ for $\alpom(0)$ are typical 
of soft hadron-hadron interactions.
As shown in Fig.~\ref{fig:Regge}~(right), the slope of the effective trajectory for VM production, 
including \rh\ photoproduction~\cite{h1-rho-photoprod},
is smaller than the value $0.25~\gevsqm$, typical for hadronic interactions.
For DVCS $\alp = 0.03 \pm 0.09 \pm 0.11~\gevsqm$~\cite{h1-dvcs}; 
for \jpsi\ photoproduction at high \modt, combining H1~\cite{h1-jpsi-large-t}
and ZEUS~\cite{z-jpsi-large-t}
measurements, $\alp = -0.02 \pm 0.01 \pm 0.01~\gevsqm$.

\paragraph {{\bf Remarks on the interaction scales}}
%

The energy dependence of the total $ep$ cross section at fixed values of \qsq\ can be 
parameterised as $F_2 \propto x^{- \lambda}$, with values of $\lambda$ increasing with \qsq,
a feature which is attributed to the increase with \qsq\ of the parton density at small $x$.
The prediction that for VM production $\delta = 2 \lambda$, when taken at the same scale, 
can thus provide information on the relevant effective scale for the reaction.
The present results clearly indicate that the variable \scaleqsqplmsq\ is a better candidate
than \qsq\ for such a unified scale, but high precision measurements of the energy 
dependence of \rh\ and \jpsi\ electroproduction remain necessary to settle the scale 
issue~\cite{levy}.

For the DVCS process, where both LO and NLO (dipole-type) diagrams contribute, the 
present high energy data seem to favour an effective scale $\approx \qsq$ rather than 
$\approx \qsq/4$ in order to ensure diffraction universality, but here also more precise data 
are required.

\section {Spin dynamics}
The VM production and decay angular distributions allow the measurement of
fifteen spin density matrix elements, which are bilinear combinations of helicity
amplitudes.
Under natural parity exchange, five $T_{\lambda_V \lambda_{\gamma}}$ amplitudes 
are independent: two $s$-channel helicity conserving (SCHC) amplitudes 
($T_{00}$ and $T_{11}$), 
two single helicity flip amplitudes ($T_{01}$ and $T_{10}$) 
and one double flip amplitude ($ T_{-11}$).

The \qsq\ dependence of the matrix elements for \rh\ and \ph\ production 
indicates that the five elements which contain products of the SCHC amplitudes 
are non-zero, whereas those formed with the helicity violating amplitudes are
generally consistent with 0.
A notable exception is the element \rczz, which
involves the product of the dominant $T_{00}$ SCHC 
amplitude with $T_{01}$,
which describes the transition from a transversely polarised photon to a longitudinal
\rh\ meson~\cite{z-rho,h1-rho-hera1}.

\begin{figure}[htbp]
\begin{center}
\setlength{\unitlength}{1.0cm}
\begin{picture}(14.8,3.5)   
\put(0.0,0.){\epsfig{file=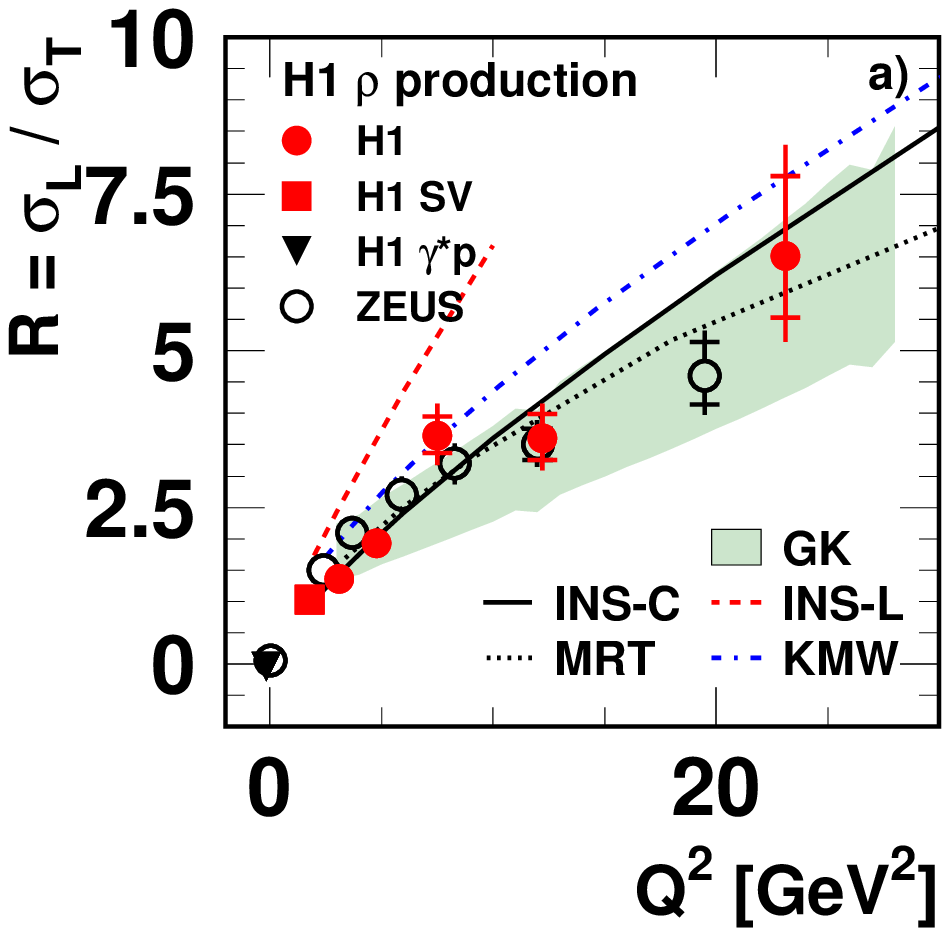,width=3.5cm}}
\put(3.7,0.){\epsfig{file=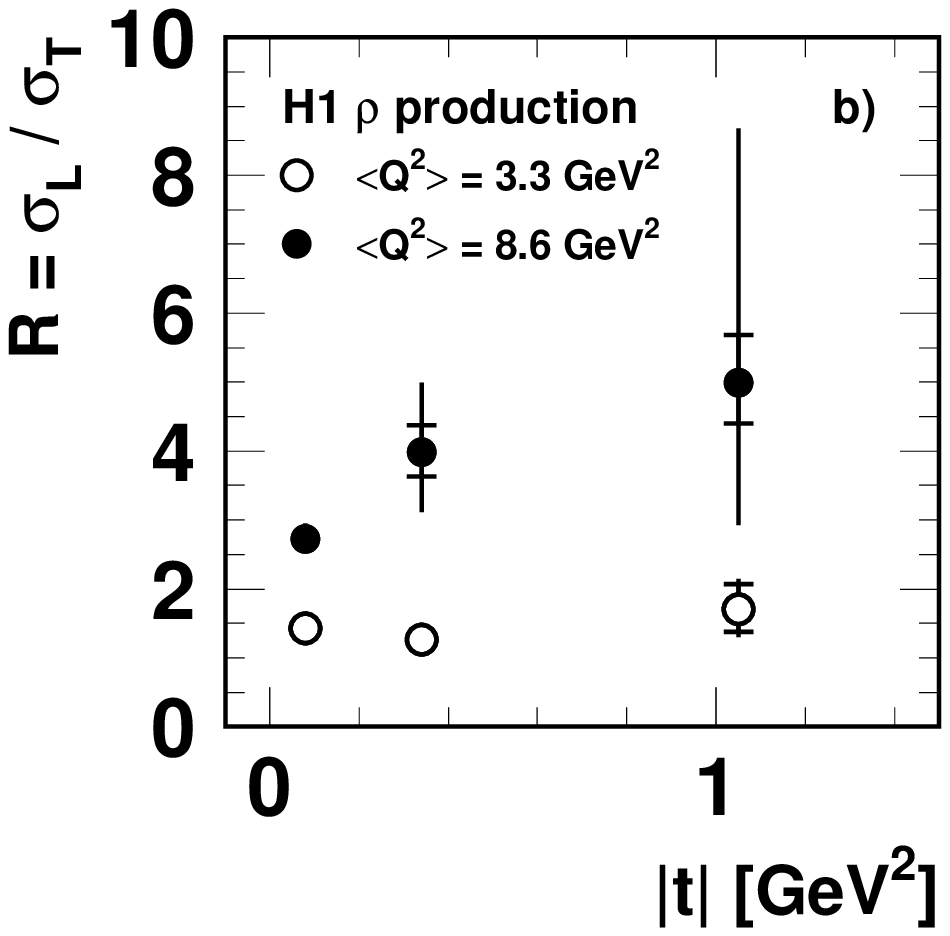,width=3.5cm}}
\put(7.5,0.2){\epsfig{file=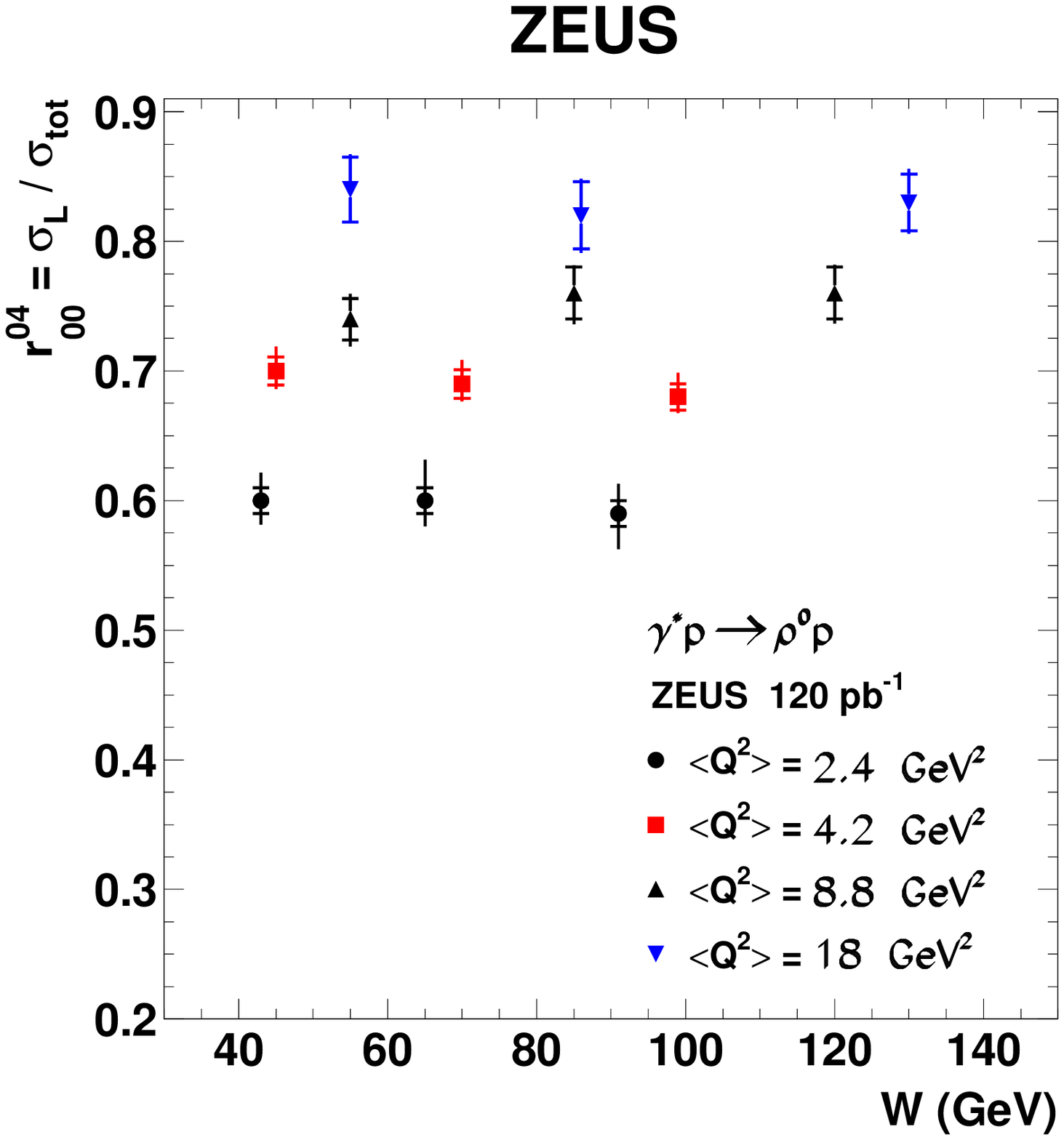,width=3.3cm}}
\put(11.1,0.2){\epsfig{file=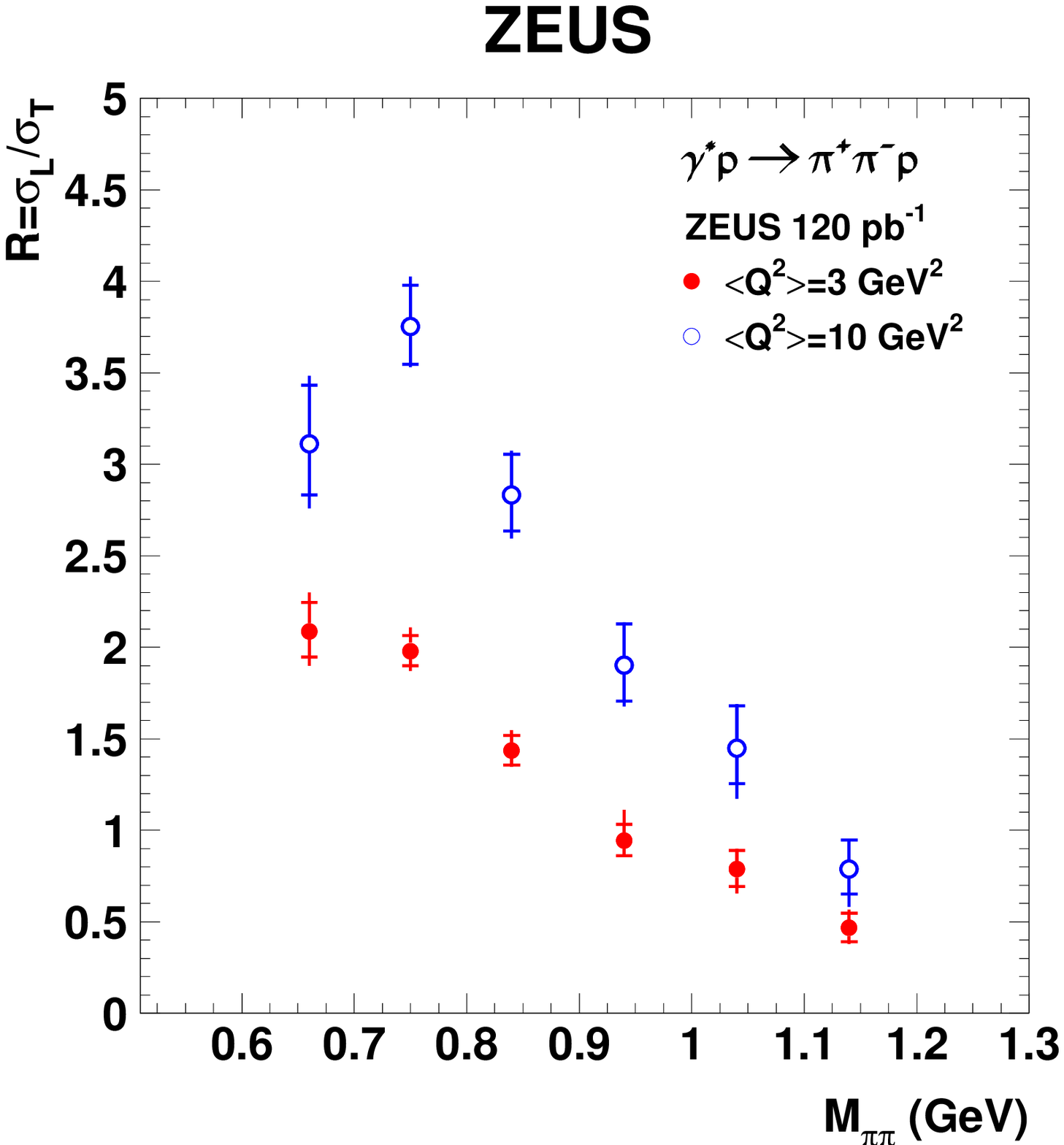,width=3.3cm}}
\end{picture}
\vspace{-0.3cm}
\caption{Measurement of $R = \sigma_L / \sigma_T$, as a function of 
\qsq\ and
\modt~\protect\cite{h1-rho-hera1},
$W$, 
and the invariant mass, $M_{\pi\pi}$, of the two decay pions~\protect\cite{z-rho}, 
for \rh\ electroproduction.}
\label{fig:R}
\end{center}
\vspace{-0.3cm}
\end{figure}
The ratio $R = \sigma_L / \sigma_T$ of the longitudinal to transverse cross sections
for \rh\ production is shown in Fig.~\ref{fig:R} as a function of \qsq, $W$, $t$, and the 
invariant mass of the two decay pions, for several domains in \qsq.

A strong increase of $R$ with \qsq\ is observed, which is tamed at large \qsq. 
These features are relatively well described by GPD and dipole models.
The \qsq\ dependence of $R$ for \rh, \ph\  and \jpsi\ production
follows a universal trend when plotted as a function of 
$Q^2  / M_{V}^2$~\cite{h1-rho-hera1}.
With the present data, no $W$ dependence is observed, but it should be stressed that
the lever arms in $W$ for fixed \qsq\ values are relatively limited.

No \modt\ dependence is observed for $R$ by ZEUS with $\modt \leq 1~\gevsq$~\cite{z-rho},
whereas an increase of $R$ with \modt\ is observed by H1 for $\qsq > 5~\gevsq$, 
$\modt \leq 3~\gevsq$~\cite{h1-rho-hera1}.
This increase can be translated into a measurement of the 
difference between the longitudinal and transverse $t$ slopes, through the relation
$R(t) = \sigma_L(t) / \sigma_T(t) \propto e^{- (b_L - b_T) |t|}$.
A slight indication ($1.5 \sigma$) is thus found for a negative value of $b_L - b_T$
($-0.65 \pm 0.14_{-0.51}^{+0.41}$), suggesting that the
average transverse size of dipoles for transverse amplitudes is larger than for
longitudinal amplitudes

The strong dependence of $R$ with the dipion mass, observed by both
experiments~\cite{z-rho,h1-rho-hera1}, cannot be attributed solely to the interference of 
resonant \rh\ and non-resonant $\pi \pi$ production, and 
indicates that the spin dynamics of \rh\ production depends of the effective
$q \bar q$ mass.    

\begin{figure}[htbp]
\begin{center}
\includegraphics[width=0.19\columnwidth]{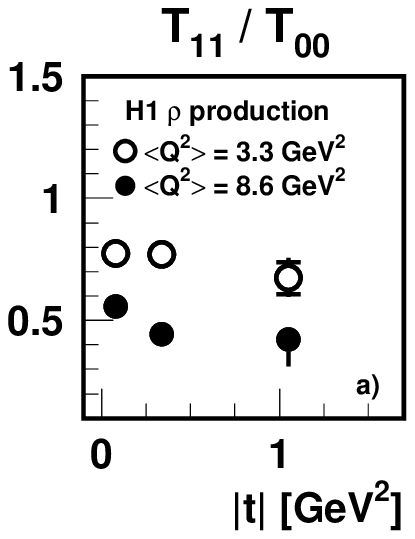}
\hspace {-0.3cm}
\includegraphics[width=0.19\columnwidth]{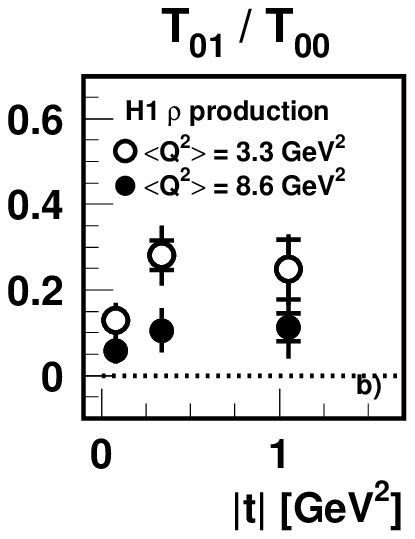}
\hspace {-0.3cm}
\includegraphics[width=0.19\columnwidth]{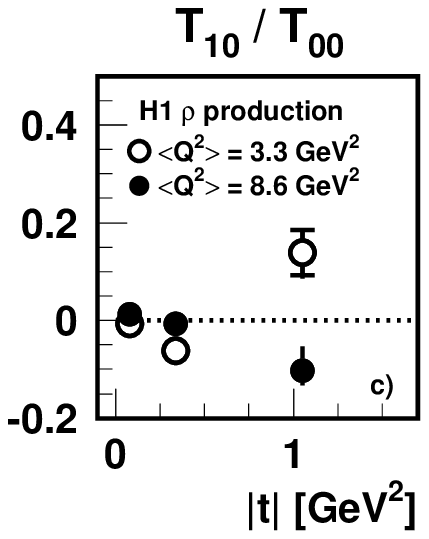}
\hspace {-0.3cm}
\includegraphics[width=0.19\columnwidth]{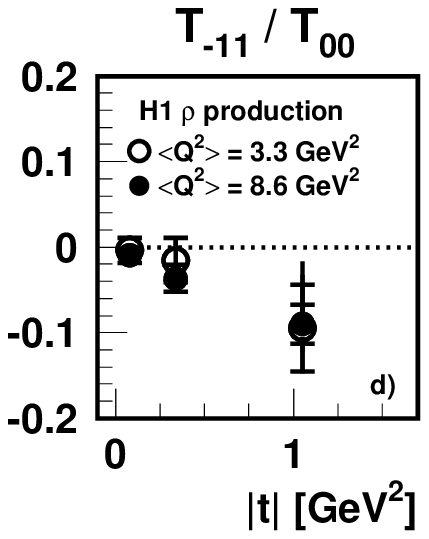}
\hspace {0.5cm}
\includegraphics[width=0.19\columnwidth]{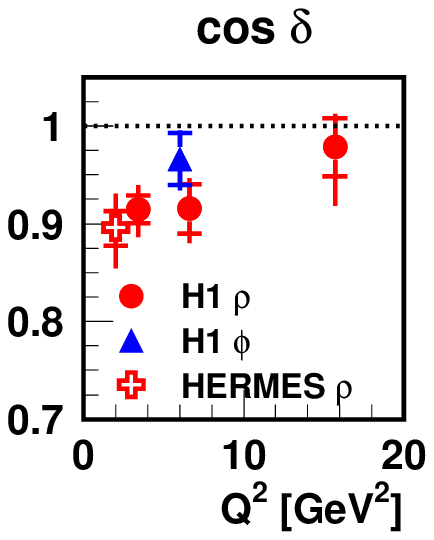}
\vspace{-0.3cm}
\caption{(a-d) Helicity amplitude ratios, as a function of $t$; (right plot)
phase difference between the two SCHC amplitudes, $T_{00}$ and 
$T_{11}$~\protect\cite{h1-rho-hera1}.}
\label{fig:ampl_ratios_t}
\end{center}
\vspace{-0.3cm}
\end{figure}

Helicity amplitude ratios are measured, under the approximation that they are
in phase, through fits to the 15 matrix elements.
The four ratios to the dominant $T_{00}$ amplitude are presented in 
Fig.~\ref{fig:ampl_ratios_t} as a function of $t$, for two domains in \qsq.
At large \qsq, a $t$ dependence compatible with the expected $\sqrt{ \modt }$
law is observed for both single helicity flip amplitudes.
A significant double-flip amplitude $T_{-11}$ is observed, which may be related
to gluon polarisation in the proton.
The $t$ dependence of $T_{11}/T_{00}$ at large \qsq, a $3 \sigma$ effect, is related to the
$t$ dependence of $R$ and supports the indication of a difference between 
the transverse sizes of dipoles in transversely and longitudinally polarised photons.

A small non-zero phase difference between the two SCHC amplitudes, which decreases with
increasing \qsq, is visible in Fig.~\ref{fig:ampl_ratios_t}.
Through dispersion relations, this non-zero value is suggestive of different $W$ 
dependences of the longitudinal and transverse amplitudes. 

\section {Large \boldmath {\modt} VM production}

In exclusive real photon and VM production at high energy and 
large \modt, 
a hard scale is present at both ends of the gluon ladder which extends over a large
rapidity range, between the struck parton in the proton (mostly gluons at small $x$) and 
the quark or antiquark from the photon fluctuation.
These processes thus offer a unique testing ground for the BFKL evolution, since no strong
$k_T$ ordering along the ladder is expected.
This is at variance with high \qsq\ VM production at low \modt, where a large scale is present
at the photon end of the ladder and a small scale at the proton end, a configuration which
is described by the DGLAP evolution. 
For real photon and \jpsi\ production, there is little uncertainty related to
the wave functions.

\begin{figure}[htbp]
\begin{center}
\setlength{\unitlength}{1.0cm}
\begin{picture}(14.,4.0)   
\put(0.2,0.1){\epsfig{file=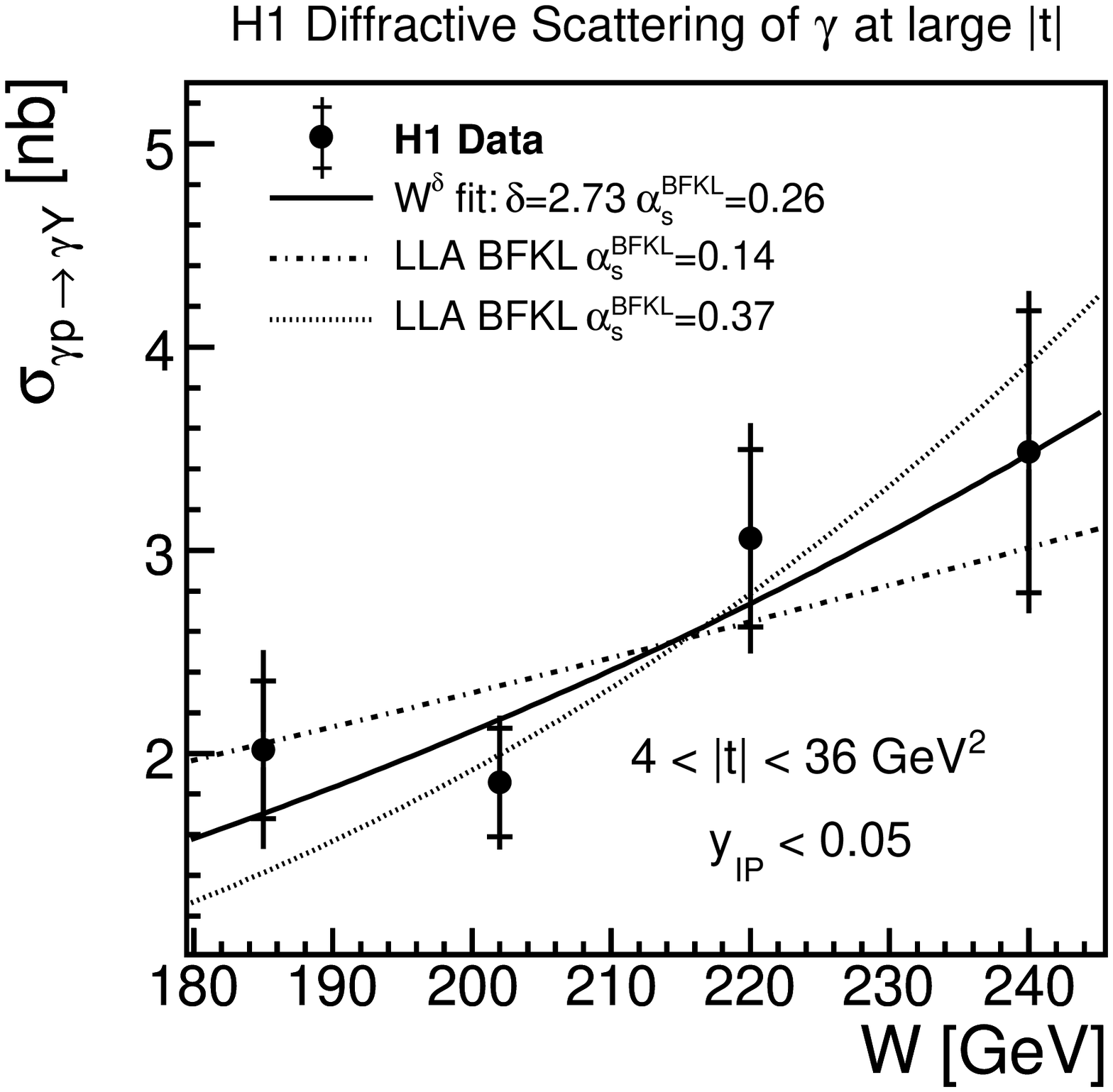,height=3.6cm,width=3.6cm}}
\put(4.1,-0.3){\epsfig{file=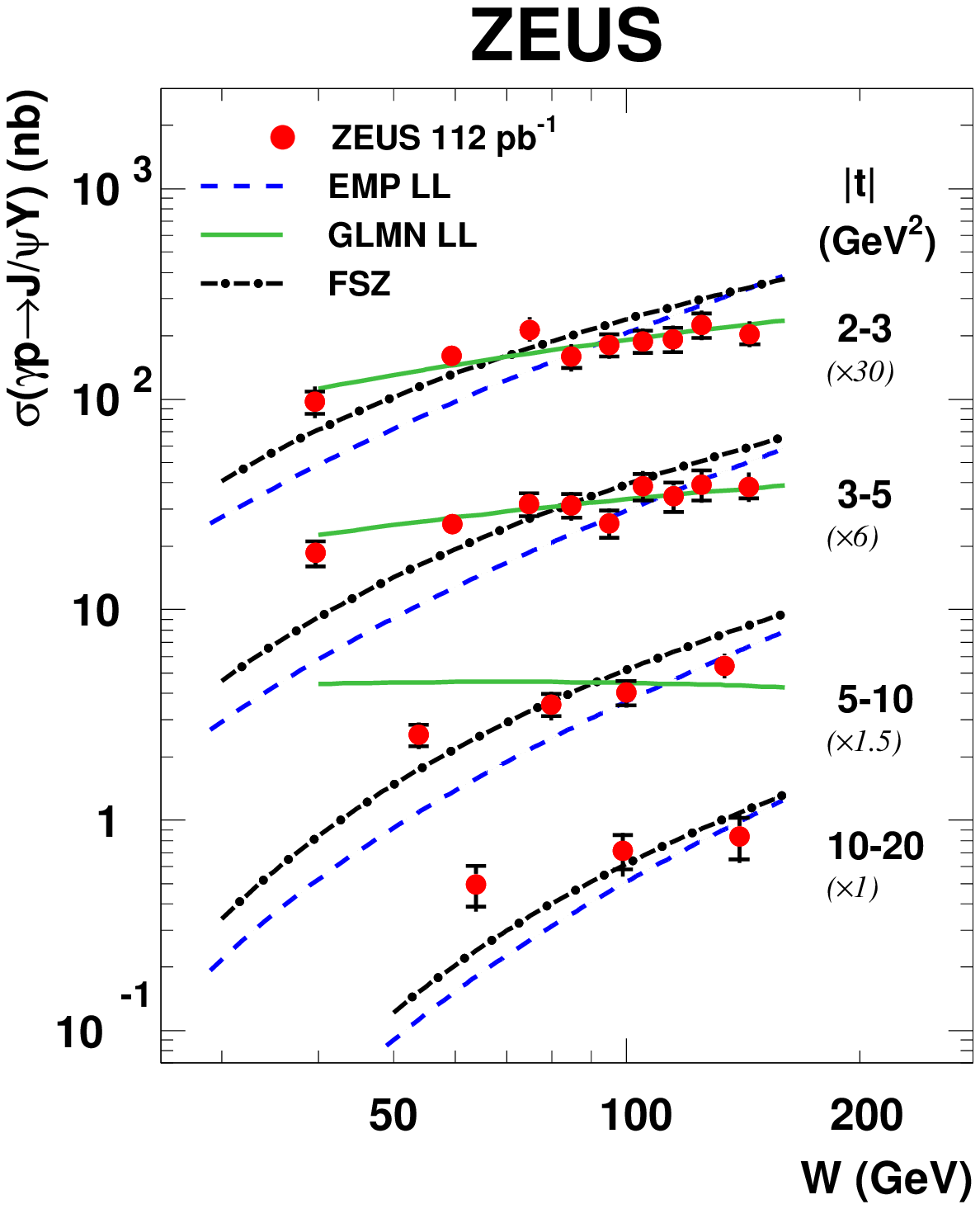,height=4.5cm,width=5.0cm}}
\put(9.3,3.4){\epsfig{file=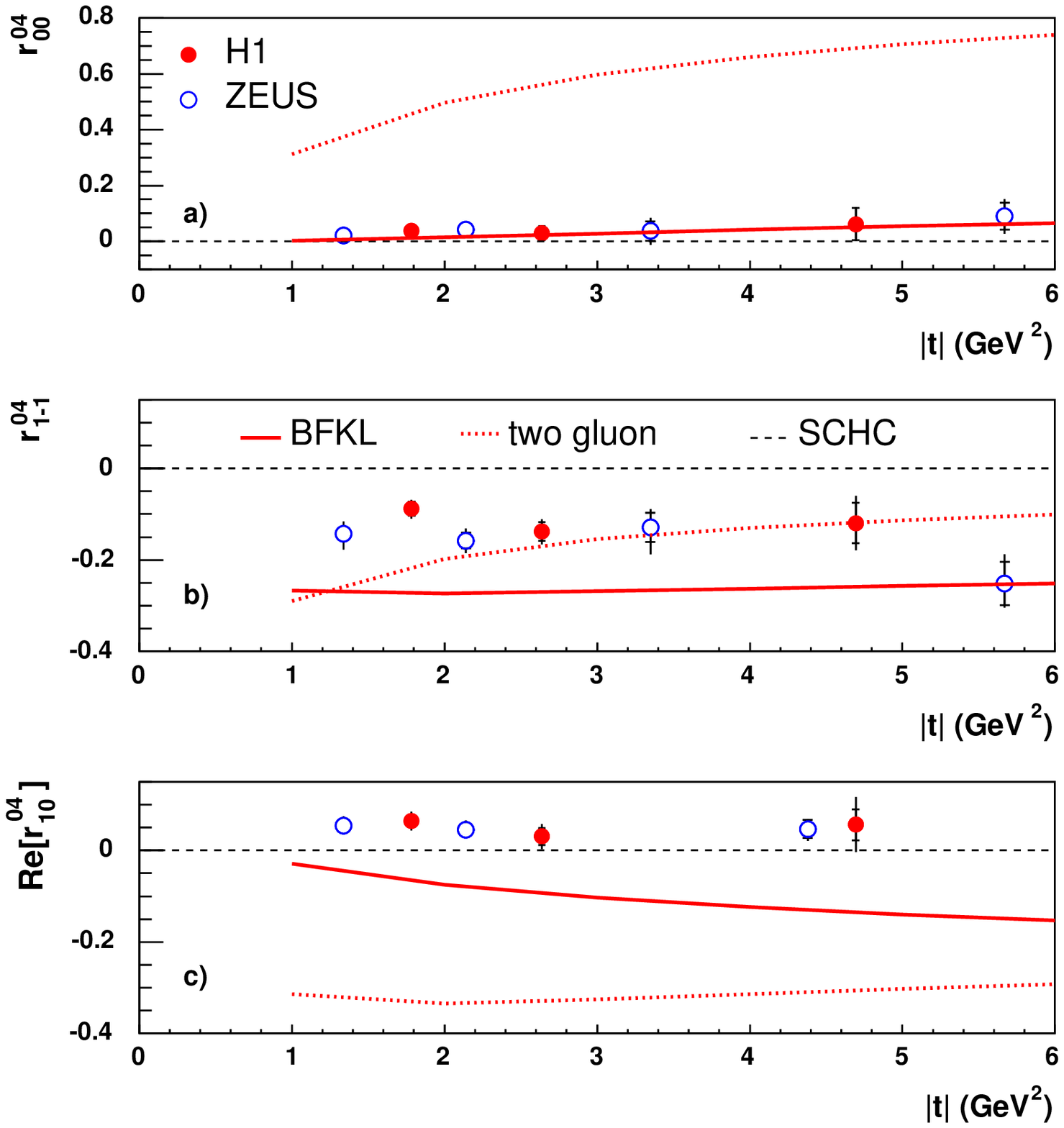,height=4.3cm,width=3.cm,angle=270}}
\end{picture}
\vspace{-0.3cm}
\caption{Large \modt\ photoproduction measurements: 
$W$ dependence of real photon~\protect\cite{h1-gamma-larget} (left) 
and \jpsi~\protect\cite{z-jpsi-large-t} production (centre); 
spin density matrix elements for \rh\ 
production~\protect\cite{h1-rho-photoprod-large-t}.}
\label{fig:hight-t}
\end{center}
\vspace{-0.3cm}
\end{figure}
A specific QCD prediction for large \modt\ production is the power-law dependence of the
\modt\ distribution, at variance with the exponential dependence for $|t| \ \lapprox\ $ 
a few~\gevsq.
The $t$ dependences for $\modt \geq 4~\gevsq$ of $\gamma$ and \jpsi\ production 
are indeed well described by power laws with exponents 
$n = 2.60 \pm 0.19 ^{+0.03}_{-0.08}$~\cite{h1-gamma-larget},
and $n = 3.0 \pm 0.1$~\cite{z-jpsi-large-t}, respectively.

Figures~\ref{fig:hight-t} (left) and (centre) present the $W$ evolutions of high \modt\ 
real photon and \jpsi\ production, respectively.
A strong $W$ dependence is observed, compatible with calculations based on
the BFKL approach, whereas the DGLAP evolution (valid for $\modt \leq m^2_{\psi}$),
predicts a significantly weaker dependence.  

The spin density matrix elements for \jpsi\ production are in agreement with 
SCHC~\cite{h1-jpsi-large-t,z-jpsi-large-t},
whereas substantial helicity flip contributions are observed in Fig.~\ref{fig:hight-t} (right)
for \rh\ production with 
$1.5\ \lsim\ \modt\ \lsim 10~\gevsq$~\cite{z-high-t,h1-rho-photoprod-large-t}, 
which can be understood in a BFKL approach with a chiral-odd component of the 
photon wave function.

\section*{Acknowledgments}
It is a pleasure to thank numerous colleagues from the ZEUS and H1 collaborations 
as well as theorists for enlightening discussions, 
and the workshop organisers for the lively discussions and the pleasant atmosphere of the 
meeting.

\begin{footnotesize}

\end{footnotesize}

\end{document}